\documentclass[letters]{aps2007}
\usepackage{graphicx}

\title{Design of a low noise, wide band, active dipole antenna for a cosmic ray radiodetection experiment}

\author[org1]{D.Charrier*}
\author[org2]{the CODALEMA collaboration}

\address[org1]{SUBATECH,  IN2P3-CNRS, Universit\'e de Nantes, Ecole
                         des Mines de Nantes;
4 rue Alfred Kastler, BP20722, 44307 Nantes, France \\ E-mail:  didier.charrier@subatech.in2p3.fr}
 
\address[org2]{LESIA, Observatoire de Paris-Meudon - Station de Radioastronomie de Nan\c{c}ay  - LAL, IN2P3-CNRS/Universit\'e de Paris Sud Orsay  -  LPSC, IN2P3-CNRS/UJF/INPG Grenoble  - ESEO, Angers  - LAOB, INSU-CNRS Besan\c con  - LPCE, SDU-CNRS Universit\'e d'Orl\'eans, France; http://codalema.in2p3.fr}

\begin{document}
\maketitleblock  
\section*{Introduction}
An active dipole antenna has been designed to measure transient electric field
induced by ultra high energy cosmic rays for the CODALEMA experiment [1,2]. 
The main requirements for this detector, composed of a low noise
preamplifier placed close to a dipole antenna, are a wide bandwidth ranging from 100~kHz to
100~MHz and a good sensitivity on the whole spectrum [3].

\section*{The active antenna concept}
\begin{figure}[h]
\begin{center}
\includegraphics[height=2.5cm]{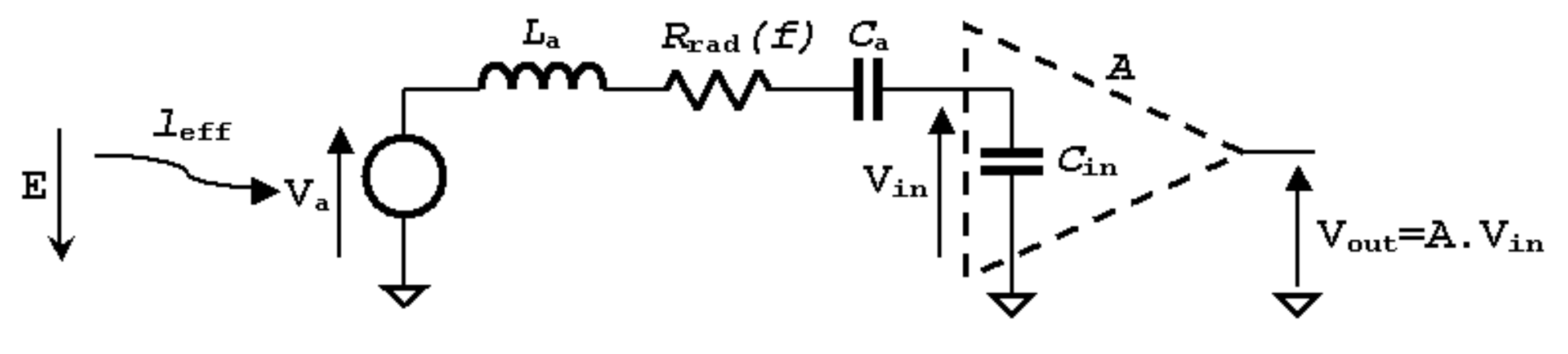}
\caption{\it Electrical equivalent model of the active antenna.}
\label{fig:shem}
\end{center}
\end{figure}

A simplified electrical model of a dipole antenna is a voltage source $V_{\mathrm{a}}$ in serial with the antenna impedance $Z_{\mathrm{a}}$, composed of a capacitance
$C_{\mathrm{a}}$, an inductance $L_{\mathrm{a}}$ and a radiation
resistance $R_{\mathrm{rad}}$. For frequencies well
below resonance (up to 1/5 of the resonance frequency $f_{\mathrm{0}}$), $Z_{\mathrm{a}}$ becomes equivalent to a capacitance due to a drop of the
radiation resistance when the frequency decreases. Two options are possible to create the active dipole
antenna: if the antenna is loaded by a preamplifier whose input impedance
is a capacitance $C_{\mathrm{in}}$ (high input impedance), then a
capacitive attenuator is obtained (see
Fig.~\ref{fig:shem}); if the antenna is loaded by a capacitive feedback
preamplifier (low input impedance), then the transfer function is given by the ratio of the antenna capacitance on the
feedback one. In both cases, the relationship between the voltage
induced on the antenna $V_{\mathrm{a}}$ and  the preamplifier output voltage
$V_{\mathrm{out}}$ becomes independent of the frequency. The first
solution was implemented.

In this scheme, the larger the antenna capacitance, the smaller the
capacitive attenuation. One way to increase the antenna capacitance 
without decreasing $f_{\mathrm{0}}$ is to enlarge the antenna wire (fat dipole). In
this configuration, the decrease of the antenna inductance lowers the
Q-factor. Consequently, the antenna can be used nearer from its resonance
frequency since its $V_{\mathrm{out}}/V_{\mathrm{a}}$ transfer function is flatter. 
Moreover the ohmic loss decreases when compared to a thin dipole.

\section*{The antenna gain}
The main advantage of a dipole antenna is an almost constant
directivity for a very wide frequency band, providing
that the antenna height is approximately less than two third of the
shortest wavelength. A constant antenna gain~$G(\theta,\varphi,f)$
assumes a lossless radiator and a perfect ground plane (infinite
conductivity). A dipole antenna is also easy to build due to its
simplicity: it is composed of two aluminium slats, each
one sizing 0.6~m length by 0.1~m width, spaced by a gap of 10~mm and
hold horizontally 1~m above the ground by a plastic mast. Assuming a perfect ground, simulations predict around $65^{\circ}$ and $90^{\circ}$ for the half power beam width
of the E and H-plane respectively.
The zenith gain is a roughly constant value of 8.5~dBi from 100~kHz to 50~MHz and decreases to
5.5~dBi at 100~MHz.

\section*{The active antenna frequency response}
The relationship between the preamplifier output
voltage~$V_{\mathrm{out}}$ and a received electric field~$E$ coming
from the direction $(\theta,\varphi)$ and parallel to the antenna, where $l_{\mathrm{eff}}$ is the antenna
effective length[4] and $A$ is the preamplifier voltage gain, is given
by:
\begin{equation} \label{eq:leffglob}
\frac{V_{\mathrm{out}}}{E}=l_{\mathrm{eff}}\frac{V_{\mathrm{out}}}{V_{\mathrm{a}}}=
\frac{c}{f}
\sqrt{\frac{R_\mathrm{rad}(f) G(\theta,\varphi,f)}{120 \pi^2}}
\times
\frac{A}{1+j2\pi Z_{\mathrm{a}}(f)C_{\mathrm{in}}f}
\end{equation}
From the lowest frequencies up to $f_0/5$, the antenna can
be considered as a short dipole without end-loading. Thus, the
radiation resistance is 
$R_{\mathrm{rad}}=197(Lf/c)^2$ with $L$,
the total antenna length [4]. Since $Z_{\mathrm{a}}$ becomes equivalent
to its capacitance $C_\mathrm{a}$, Eq.~\ref{eq:leffglob} leads to:
\begin{equation} \label{eq:leffbf}
\frac{V_{\mathrm{out}}}{E}\simeq\frac{L}{\sqrt{6}}
\sqrt{G(\theta,\varphi)}
\times
\frac{A}{1+\frac{Cin}{Ca}}
\end{equation}
Assuming a perfect ground plane Eq.~\ref{eq:leffbf} implies that $V_{\mathrm{out}}/E$ is constant from
100~kHz to approximately $f_0/5$ in a given direction. At the zenith, effective lengths of
$0.493\sqrt{G(\theta,\varphi)}=$1.31~m for the low frequencies, and
$0.628\sqrt{G(\theta,\varphi)}=$1.19~m at 100~MHz, can be deduced.

The knowledge of the antenna impedance is very important. It has
been measured with a vector network analyser supplying power to the antenna radiator through a 15~m cable and a balun RF
transformer. The measured antenna capacitance (including parasitic capacitance) is 10~pF at 10~MHz
and the resonance frequency is 112~MHz. With these values of
$Z_{\mathrm{a}}$, it becomes possible to calculate between 40~MHz and 170~MHz the $V_{\mathrm{out}}/V_{\mathrm{a}}$ transfer
function of the antenna with its
preamplifier. A simulation with accurate values of $R_{\mathrm{rad}}$ from 100~kHz to 100~MHz is under study to plot the
overall active antenna frequency response $V_{\mathrm{out}}/E$.

\section*{The preamplifier}
\begin{figure}[h]
\begin{center}
\includegraphics[width=\linewidth,height=5.5cm]{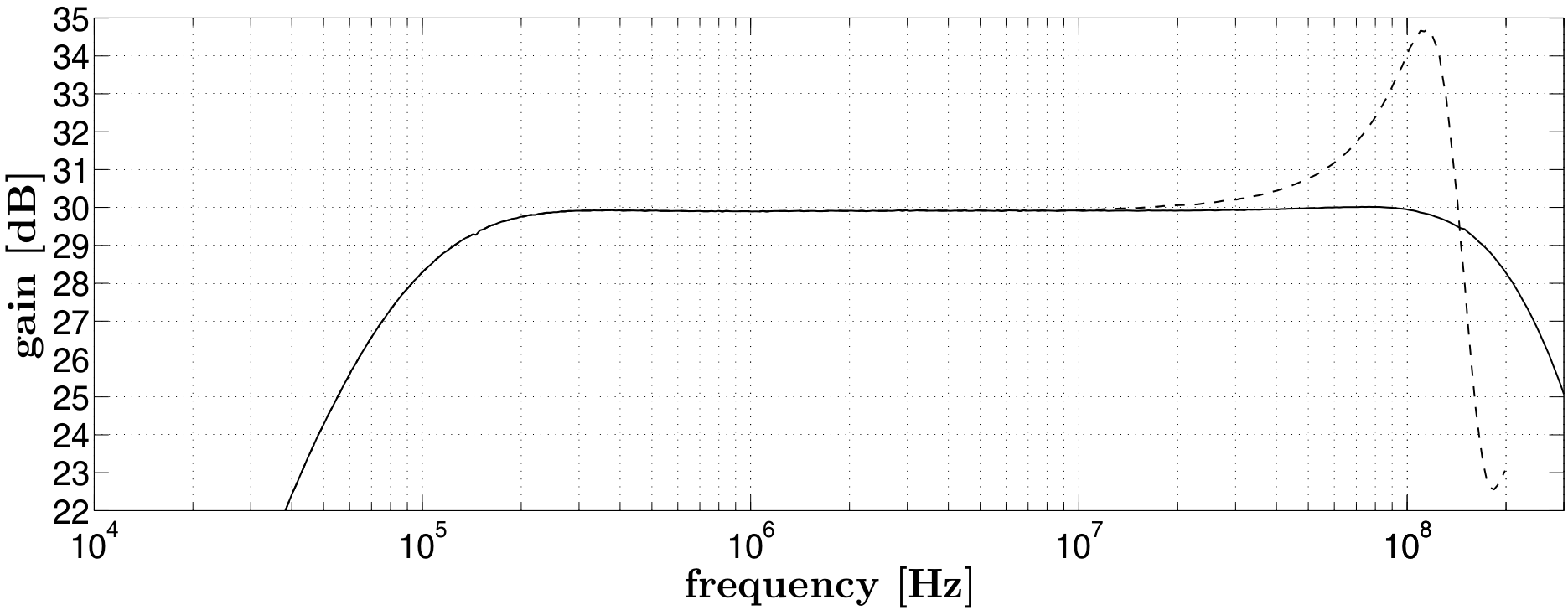}
\caption{\it Preamplifier gain measurements
  ($V_{\mathrm{out}}/V_{\mathrm{a}}$): the solid line is the measured
  gain replacing $Z_{\mathrm{a}}$ by a 10pF dummy antenna
  capacitance, whereas the dotted line is the gain calculated by taking
  into account the measured values of $Z_{\mathrm{a}}$.}
\label{fig:gain}
\end{center}
\end{figure}

To fulfil the noise and bandwidth constraints, a dedicated preamplifier was
designed using the AMS BiCMOS 0.8~$\mu$ technology. This ASIC
contains three fully differential amplifiers: the input one is low
noise with a voltage gain of 33~dB and a capacitive input impedance of 10~pF. The gain of the middle amplifier
is digitally adjustable from 9.5 to 16.8~dB. The power output amplifier is designed
to drive a 100~$\Omega$ load. The maximum input dynamics is 24~mV and the  consumption is
0.25~W. Because a low noise is required from the lowest frequencies and the antenna impedance is inversely proportional
to the frequency, a MOS transistor was chosen  due to its lack of current noise. The flicker noise is reduced choosing a P channel
whereas the thermal noise is lowered by sizing a wide PMOS
transistor. Since the widest the input transistor, the highest the input
capacitance, there is one optimal size of the input CMOS transistor
depending on the antenna capacitance. On the preamplifier output
noise density measurement shown on Fig.~\ref{fig:noise}, a dummy
impedance equivalent to $Z_{\mathrm{a}}$ is connected to the differential input. This ASIC is mounted on a small printed
board with a balun output transformer allowing to drive a 50~$\Omega$ load through a coaxial
cable. The $V_{\mathrm{out}}/V_{\mathrm{a}}$ ratio of this preamplifier board (Fig.~\ref{fig:gain}) is 30~dB with a
10~pF dummy antenna capacitance, and the -3~dB bandwidth is ranging
from 80~kHz to 230~MHz. With the measured values of $Z_{\mathrm{a}}$, it exhibits a maximum value of 34.7~dB at
113~MHz due to the antenna resonance. Two external feedback resistors connect the
differential outputs to the differential inputs to bias the
preamplifier. Moreover, thanks to $C_{\mathrm{in}}$, they act
as an active first order high pass filter whose cut off
frequency can be easily adjusted from 10~kHz to more than 1~MHz.
 
\section*{The active antenna sensitivity}
\begin{figure}[h]
\begin{center}
\includegraphics[width=\linewidth,height=6.0cm]{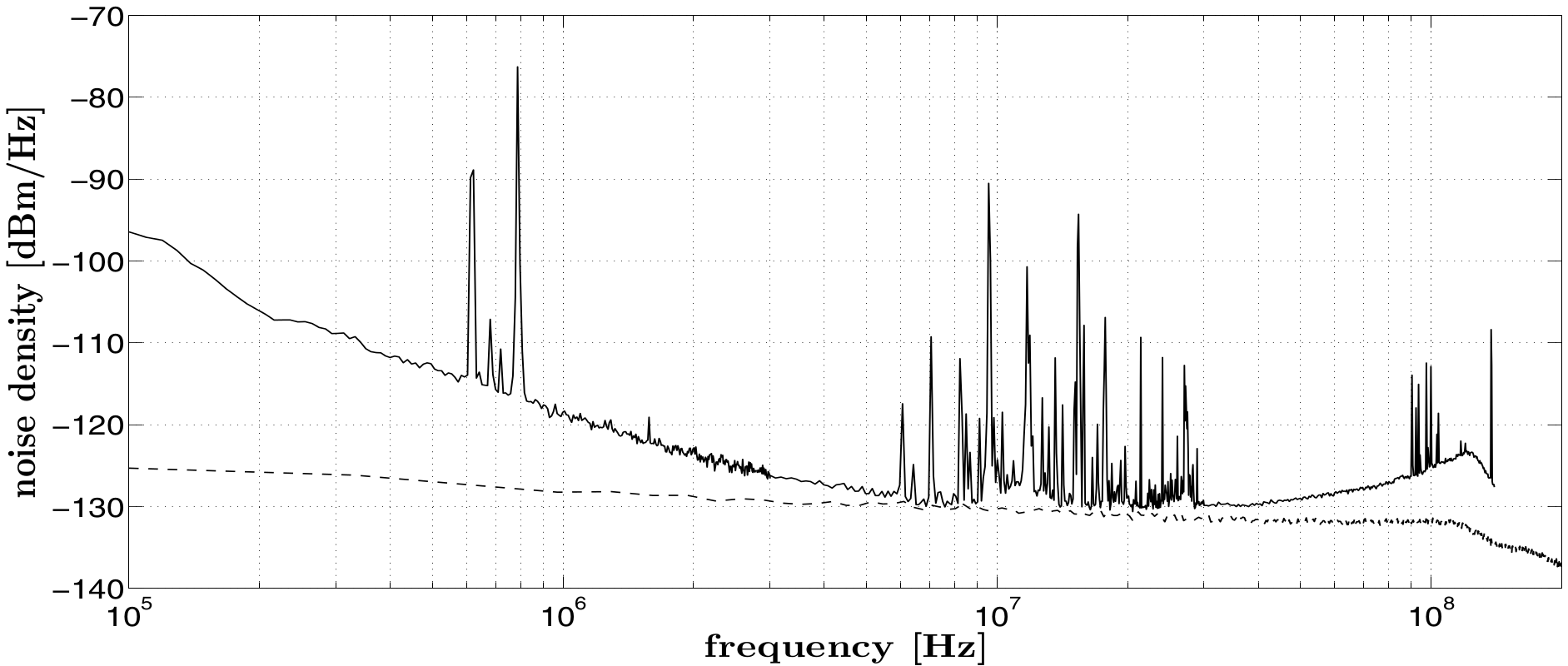}
\caption{\it Noise density measured with (solid line) the active antenna and a
  spectrum analyser at Malargue (Argentina) and without (dotted
  line) the antenna radiator.}
\label{fig:noise}
\end{center}
\end{figure}

Two noise sources should be considered to evaluate the antenna
sensitivity: the preamplifier electronic noise $v_{amp}^2$ which is mainly dominated
by the first transistor stage, and the noise resulting from the sky background temperature $T_{\mathrm{sky}}$. It generates an
equivalent noise source $v_{sky}^2=4k_{\mathrm{B}}T_{\mathrm{sky}}R_{\mathrm{rad}}$. The two noise densities shown on
Fig.~\ref{fig:noise} are measured with the active antenna
measuring the sky, and with the preamplifier whose input is connected
to a dummy antenna impedance. Besides in the 5-30~MHz range, the sky noise
floor is clearly greater than the preamplifier noise. The atmospheric
noise seems to dominate below 5~MHz  whereas the galactic noise is the
greatest above 30~MHz. The $v_{sky}^2/v_{amp}^2$ ratio characterises
the active antenna sensitivity: 0~dB at 50~MHz and 4.5~dB at 100~MHz. At 100~MHz, with a preamplifier gain of
34~dB (Fig.~\ref{fig:gain}) and an output electronic noise of
-131.7~dBm/Hz (Fig.~\ref{fig:noise}), a galactic noise of
-161.2~dBm/Hz at the preamplifier input can be deduced. With
$R_{\mathrm{rad}}=51.3~\Omega$, $T_{\mathrm{sky}}=1340K$ is calculated. The same
calculation at 50~MHz, where $R_{\mathrm{rad}}=5.3~\Omega$ gives
$T_{\mathrm{sky}}=9480K$. These two temperature estimations agree
fairly well with the known values [4]. One should note that those
noises could be used as a calibration method for the CODALEMA experiment.

\section*{Conclusion}
This active antenna gives good results for its purpose. Since June
2005, a cross shape array of 16 antennas has been installed and is
currently under operation at the Nan\c{c}ay radio-observatory.
A new ASIC design with an up-to-date technology is under work from
which an electronic noise improvement of 3~dB is expected.

\section*{References}
[1] D. Ardouin {\it{et al.}}, Nucl. Instruments and Methods, A 555 (2005) 148-163. 

[2] D. Ardouin {\it{et al.}}, Astroparticle Physics, 26 (2006) 341-350.

[3] D. Ardouin {\it{et al.}}, Proc. X Pisa 2006 Meeting,
Nucl. Instrum. Meth. A in press.

[4] J.D. Kraus, Antennas, McGraw-Hill, New York, 1988.
 
\end{document}